\documentclass[a4paper,onecolumn,draftclsnofoot,11pt]{IEEEtran}

\usepackage[cmex10]{amsmath} 
\usepackage{amssymb,amsthm,bm}
\usepackage{mathtools}
\usepackage{bm}

\usepackage{tikz}
\usetikzlibrary{calc,positioning,fit,shapes}
\usepackage{pgfplots}
\pgfplotsset{compat=1.9}

\usepackage{calc}
\usepackage{cite}

\newtheorem{theorem}{Theorem}
\newtheorem{lemma}{Lemma}

\usepackage{bbm}

\newcommand{\EE}{\mathbb{E}}

\IEEEoverridecommandlockouts
\begin{document}

\title{Soft-TTL: Time-Varying Fractional Caching}

\author{
Jasper Goseling and Osvaldo Simeone
\thanks{
The work of O. Simeone was partially
	supported by the European Research Council
	(ERC) under the European Union Horizon 2020 research and innovation program (grant agreements 725731) and by the U.S. NSF through grant CCF-1525629.

J. Goseling is with the Department of Applied Mathematics, University of Twente, the Netherlands. (email: j.goseling@utwente.nl)

O. Simeone is with Centre for Telecommunications Research, Department of
	Informatics, King'€™s College London, UK. (email: \mbox{osvaldo.simeone@kcl.ac.uk}) }
}

\maketitle

%
%
%
\begin{abstract}
Standard Time-to-Live (TTL) cache management prescribes the storage of entire files, or possibly fractions thereof, for a given amount of time after a request. As a generalization of this approach, this work proposes the storage of a time-varying, diminishing, fraction of a requested file. Accordingly, the cache progressively evicts parts of the file over an interval of time following a request. The strategy, which is referred to as \emph{soft-TTL}, is justified by the fact that traffic traces are often characterized by arrival processes that display a decreasing, but non-negligible, probability of observing a request as the time elapsed since the last request increases. An optimization-based analysis of soft-TTL is presented, demonstrating the important role played by the \emph{hazard function} of the inter-arrival request process, which measures the likelihood of observing a request as a function of the time since the most recent request.
\end{abstract}


%
%
\section{Introduction} \label{sec:intro}
Caching of popular files is a key enabling technology for content delivery networks and is finding applications also at the wireless edge (see, e.g., \cite{wang2015optimal,sengupta2016cloud}). Conventionally, an entire file is stored in a cache when a previously uncached content is requested by some user. Furthermore, an increasingly popular approach for managing the cache memory prescribes that a file be retained in the cache for a fixed amount of time, known as Time-to-Live (TTL), which is generally to be optimized based on the statistics of the users' requests~\cite{dehghan2016utility}. By properly adjusting the parameters in a TTL cache, it has been shown that TTL caching can mimic the behavior of traditional caching mechanisms such as Least Recently Used (LRU) and Least Frequently Used (LFU)~\cite{fofack2012analysis}.   

In contrast to the classical design of caching entire files, it is well understood that there are potential advantages to be accrued by storing only fractions of popular files: (\emph{i}) Users may only consume part of the content, e.g., watch only the first few minutes of a movie~\cite{wang2015optimal,yang2017audience,altman2014distributed}; and (\emph{ii}) In streaming applications, the cached fraction can be used to fill the buffers and immediately start playback, while the rest of the file is downloaded from the library~\cite{leconte2016placing,sengupta2016cloud}. Under these conditions, caching a fraction of a file helps freeing up space to store more files, hence sharing the benefits of caching across a larger range of popular contents.

\begin{figure}
	\centering
	\begin{tikzpicture}
	\begin{axis}[
	width=12cm,
	height=7cm,
	xlabel=time,ylabel=cached fraction, 
	font=\scriptsize,
	xtick=\empty,
	extra x ticks = {0, 2.3, 5.5, 7.2},
	extra x tick style = {grid=major},
	extra x tick labels = {$\tau_i(n-1)$, $t$, $\tau_i(n)$, $\tau_i(n+1)$},
	ytick=\empty,,
	extra y ticks = {0,2,3,4,6},
	extra y tick labels = {$\mu_{i,4}$, $\mu_{i,3}$, $\mu_{i,2}$, $\mu_{i,1}$, $\mu_{i,0}$}
	]
	
	\addplot[
	line width=.3mm,color=blue, 
	const plot mark left,
	]
	coordinates {
		(0,6) (1,4) (2,3) (3,2) (4,0) (5.5,6) (6.5,4) (7.2,6) (8.2,4) (9.2,3) 
	};
	
	\draw[latex-latex] (axis cs: 0,5) -- node[below] {$T$} (axis cs: 1,5);
	\draw[latex-latex] (axis cs: 0,1.5) -- node[below] {$t - \tau_i(N_i(t))$} (axis cs: 2.3,1.5);
	
	\end{axis}
	\end{tikzpicture}
	\caption{Soft-TTL via time-varying fractional caching: A file $i$ requested a times $\tau_i(n)$ for $n=1,2,...$, and a decreasing fraction of the file is stored in the cache for a subsequent interval of time.}
	\label{fig:overview}
\end{figure}
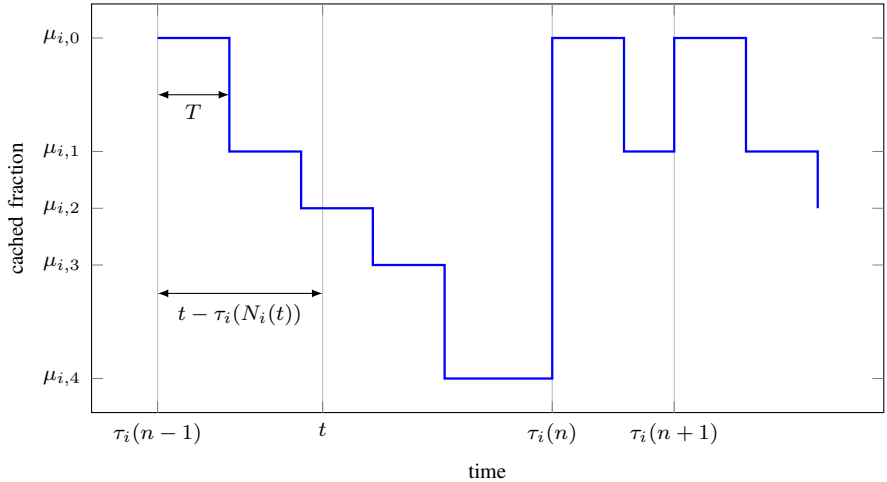

In this paper, we generalize the idea of caching fractions of files by enabling the storage of a time-varying, diminishing, fraction of a requested file. This is illustrated in Fig. \ref{fig:overview}, in which, upon receiving a request at times $\tau_i(n)$ for $n=1,2,...$, a fraction of a previously uncached file $i$ is stored, and the cache progressively evicts parts of the file over a subsequent interval of time. The approach, which we refer to as \emph{soft-TTL}, can be seen as a generalization of the concept of TTL, whereby an entire file is cached for a given amount of time after a request, as well as of \emph{fractional-TTL}, in which a fraction of a file is stored over some time interval following a request. 

Soft-TTL is justified by the fact that traffic traces are often characterized by arrival processes that display a decreasing, but non-negligible, probability of observing a request as the time elapsed since the last request increases  \cite{feldmann2000characteristics,jung2003modeling}. Extending the argument in favor of fractional caching, soft-TTL can potentially use the available cache memory more effectively by adapting to the traffic statistics even in the presence of late arrivals.

In this paper, we present an optimization-based analysis of soft-TTL by comparing it on an equal footing with conventional TTL and fractional TTL. We provide analytical results regarding the optimization of the mentioned caching policies given the traffic statistics. We specifically demonstrate the important role played by the \emph{hazard function} of the inter-arrival request process, which measures the likelihood of observing a request as a function of the time elapsed since the last request~\cite[Ch. 2]{rausand2004system}. Numerical results show the advantages offered by the additional design degrees of freedom enabled by soft-TTL.

%
%
%
\section{Model} \label{sec:model}
We consider a standard set-up in which a collection of $J$ popular files in a given content library are requested according to independent renewal processes. Each file  $i\in\{1,\dots,J\}$ has size $s_i$ in bits. As illustrated in Fig. \ref{fig:overview}, let $\tau_i(n)$ for $n=1,2,...$ and $i\in\{1,\dots,J\}$ denote the time of the $n$-th request for file $i$; and $N_i(t)$, for $t \geq 0$, the number of corresponding requests up to time $t$. For notational convenience and without loss of generality, we assume that $\tau_i(0)=0$. The time between two requests for the same file $i$ has a cumulative distribution\begin{equation}
F_i(t) = \mathrm{Pr}\left( \tau_i(n) - \tau_i(n-1) \leq t \right).
\end{equation} We assume that function $F_i(t)$ is continuous, and use $\lambda_i = \EE[X_i]^{-1}$ to denote the request rate, where $X_i$ is a random variable distributed according to $F_i(t)$. 




We consider a single cache that can store entire files or a fraction thereof. According to the principle of soft-TTL introduced in the previous section, the fraction of a file that is cached can change over time. Let $\mu_i(t)$ denote the fraction of file $i$ that is cached $t$ seconds after its last request, i.e., after the $N_i(t)$th request. Function $\mu_i:[0,\infty)\to[0,1]$ is non-increasing over time and is referred to as the \emph{caching policy} for file $i$. The constraint that the caching policy is non-increasing follows from the fact that a cache can evict part of a file from its storage, but it cannot store more than it currently has until a request for the file occurs. 

As seen in Fig. \ref{fig:overview}, we specifically restrict our attention to caching policies that take at most a finite number of values. Furthermore, changes can only occur at equally spaced intervals of time equal to $T$ for an overall amount of time equal to $KT$ for some integer $K$. Accordingly, we consider caching policies $\mu_i(t)$ that are given as 
\begin{equation}\label{eq:policy}
\mu_i(t) =
\begin{cases}
\mu_{i,0}, &\hspace{-2mm}\text{if}\ t  < T, \\
\mu_{i,k}, &\hspace{-2mm}\text{if}\ kT \leq t  < (k+1)T, \text{ for } k=1,...,K-1  \\
\mu_{i,K}, &\hspace{-2mm}\text{if}\ t \geq KT
\end{cases}
\end{equation}
for some cached fractions  $1\geq\mu_{i,0}\geq\mu_{i,1}\geq\dots\geq\mu_{i,K}\geq 0$.

Broadly speaking, the aim of caching is to minimize the cost of serving requested files. In this paper, following a standard approach (see, e.g., \cite{dehghan2016utility}), we will formulate this task as the maximization of a utility function. To this end,
for any file $i$, we define a utility function $w_i(\mu)$ that measures the utility accrued if file $i$ a requested at a time in which it is cached for a fraction $\mu$. We assume that the functions $w_i:[0,1]\to[0,\infty)$ are strictly concave and increasing. This choice reflects the scenarios listed in Sec. I in which caching yields decreasing gains as a larger fraction of a file is stored. The long-term average utility for a given file $i$ is hence given as the limit
\begin{equation}\label{eq:Wi}
W_i = \lim_{t\to\infty}\frac{1}{t}\sum_{n=1}^{N_i(t)} w_i(\mu_i(\tau_i(n)-\tau_i(n-1))).
\end{equation}
We aim at maximizing the standard $\alpha$-fair utility metric over all files~\cite{mo2000fair}, which is defined as
\begin{equation}\label{eq:alphautility}
\mathcal{W} =
(1-\alpha)^{-1}\sum_{i=1}^J W_i^{1-\alpha}, 
\end{equation}
where $\alpha\neq 1$ is a fairness parameter. Note that, at the two extremes, setting $\alpha=0$ corresponds to maximizing the sum utility, while $\alpha\to\infty$ corresponds to the strictest criterion of max-min fairness.

The utility (\ref{eq:alphautility}) will be optimized over the caching policies $\mu_i(t)$ under a long-term average cache capacity constraint, also considered in, e.g., \cite{dehghan2016utility}, which is given as
\begin{equation}\label{eq:Ci}
\sum_{i=1}^J C_i \leq C,\ \text{with }C_i = \lim_{t\to\infty}\frac{s_i}{t}\int_{0}^t \mu_i(s) ds,
\end{equation}
where $C_i$ is the long-term average cache occupancy of file $i$. As discussed in the next section, the general problem formulation described here subsumes as special cases the optimization of the conventional TTL and fractional-TTL methods.

To conclude this section, we define the following two key quantities 
\begin{align}
F_{i,k} &=F_i((k+1)T)-F_i(kT), \\
A_{i,k} &= \int_{kT}^{(k+1)T}(1-F_i(t))dt,
\end{align}
for $k=0,\dots,K-1$, as well as $F_{i,K} = 1-F(KT)$ and $A_{i,K} = \int_{kT}^{\infty}(1-F(t))dt$.

%
%
%
\section{Preliminaries: TTL and Fractional-TTL} \label{sec:preliminaries}

In this section, we formulate optimization models that provide the utility-maximizing TTL and fractional-TTL policies. A fractional-TTL policy is obtained as a special case of the general strategy (\ref{eq:policy}) by imposing that the same fraction $\nu_i$ be cached for some TTL period $LT$ defined by an integer $0\leq L\leq K$, i.e., by setting $\mu_{i,0} = \mu_{i,1} = \dots = \mu_{i,L} =\nu_j$ and $\mu_{i,L+1}=\mu_{i,L+2}=\dots=\mu_{i,K}=0$. The TTL policy further imposes the constraint that the cached fraction be equal to one, i.e., $\nu_i=1$.

To proceed, we first establish
the following general result on the utility $W_i$ and cache occupancy $C_i$ for each file $i$.
\begin{lemma} \label{lem:reward}
For $i=1,\dots,J$ we have
\begin{equation} \label{eq:reward}
W_i = \lambda_i\sum_{k=0}^K w_i(\mu_{i,k})F_{i,k},\quad
C_i = \lambda_i s_i \sum_{k=0}^K \mu_{i,k} A_{i,k},
\end{equation}
with probability $1$.
\end{lemma}
\begin{IEEEproof}
For the utility (\ref{eq:Wi}), we can write $\lambda_i\EE\left[w_i(\mu_i(X_i))\right]$, where the equality holds with probability $1$ by the fundamental theorem in renewal reward theory (see, for instance,~\cite[Theorem~3.6.1]{ross}). This is seen  by identifying $w_i(u_i(X_i))$ as the reward that is obtained if a request, i.e., a renewal, is observed after time $X_i$ from the previous. A similar argument applies for the cache occupancy $C_i$ in (\ref{eq:Ci}). To this end, we interpret the quantity
$\int_0^{X_i} \mu_i(s)ds$ 
as the reward obtained at the end of a renewal cycle. 
\end{IEEEproof}

To formulate the optimization of utility (\ref{eq:alphautility}) under the cache capacity constraints in (\ref{eq:Ci}) for fractional TTL we introduce a binary variable $\theta_{i,k}$ such that
\begin{equation}
\theta_{i,k} =
\begin{cases}
1,\quad &\text{if } k\leq L, \\
0,\quad &\text{if } k>L.
\end{cases}
\end{equation} The resulting optimization problem follows directly from Lemma 1 and is given as 
\begin{align}
\text{maximize}\ & (1-\alpha)^{-1}\sum_{i=1}^J\left(\lambda_i \sum_{k=0}^K  w_i(\mu_{i,k})F_{i,k}\right)^{1-\alpha}, \label{eq:frac_objective}
\\
\text{subject to}\ & \sum_{i=1}^J \lambda_i s_i \sum_{k=0}^K \mu_{i,k} A_{i,k} \leq C, 
\\
& 1\geq \mu_{i,0}\geq \mu_{i,1}\geq\dots\geq \mu_{i,K} \geq 0, \\ 
& 0 \leq \nu_i \leq 1, \label{eq:frac_constraintA} \\
& \theta_{i,k} \in \{0,1\}, \\
& -\theta_{i,k} \leq \mu_{i,k} \leq \theta_{i,k}, \\
& \theta_{i,k} - 1 \leq \mu_{i,k} - \nu_i \leq 1 - \theta_{i,k}, \label{eq:frac_constraint_last}
\end{align} where the optimization is over the variables $\{\mu_{i,k},\theta_{i,k},\nu_i\}$. The problem above is a mixed-binary convex program that can be solved by modern optimization software such as Gurobi~\cite{gurobi}. The corresponding problem for TTL is obtained by setting $\nu_i=1$.

%
%
%
\section{Soft-TTL} \label{sec:soft}
In this section, we tackle the general problem of optimizing the utility (\ref{eq:alphautility}) under cache capacity constraints (\ref{eq:Ci}) for soft-TTL policies. We first demonstrate in Theorem 1 that the problem results in a convex optimization program. We then offer in Theorem 2 an interpretation of the optimal soft-TTL policy in terms of the hazard function of the request processes.

\begin{theorem} \label{th:soft}
The problem of optimizing the utility (\ref{eq:alphautility}) under cache capacity constraints (\ref{eq:Ci}) for soft-TTL can be written as the following convex program
\begin{align}\label{eq:problemsoft}
\text{maximize}\  & (1-\alpha)^{-1}\sum_{i=1}^J\left(\lambda_i \sum_{k=0}^K w_i(\mu_{i,k})F_{i,k}\right)^{1-\alpha}, 
\\
\text{subject to}\ & \sum_{i=1}^J \lambda_i s_i \sum_{k=0}^K \mu_{i,k} A_{i,k} \leq C, 
\\
 & 1\geq\mu_{i,0}\geq\mu_{i,1}\geq\dots\geq\mu_{i,K}\geq 0. 
\end{align}
\end{theorem}
\begin{IEEEproof}
The theorem follows directly from Lemma 1 and basic properties of convex functions~\cite{boyd2004convex}.
\end{IEEEproof}

Given the convexity of the problem (\ref{eq:problemsoft}), an optimal soft-TTL policy can be found with any general-purpose convex solver. To obtain further insights into the optimal soft-TTL policy and its relationship to fractional-TTL and TTL, we now consider the special case of the problem (\ref{eq:problemsoft}) with $J=1$ file. With this simplification, the optimization can be handled in a manner similar to continuous nonlinear resource allocation problems as considered in, e.g.,~\cite{bretthauer1995nonlinear}. 

To elaborate on this point, for the rest of this section, we drop the index $i$ since we consider $J=1$ file, and we set $\lambda_1=s_1=1$ without loss of generality. The following theorem details the optimal soft-TTL solution. An interpretation of the result is discussed after the theorem. We note that the extension to multiple files would follow by properly optimizing the partition of the cache across the different files, which generally requires numerical optimization.

\begin{theorem} \label{th:analytical}
Assume that $J=1$ and that the utility function $w$ is differentiable, strictly concave, and increasing. For any $\gamma\geq 0$, let $\bar\mu_k(\gamma)$ denote the solution\footnote{
For completeness, if $ \gamma\frac{A_k}{F_k}$ is above (or below) the range of values for $w'$, then the quantity $\bar\mu_{k}(\gamma)$ should be set to $0$ (or $1$).
}
of the equality
\begin{equation} \label{eq:barmusol}
 w'(\bar\mu_{k}(\gamma)) = \gamma\frac{A_k}{F_k}
\end{equation}
and let
\begin{equation}
\mu^*_{k}(\gamma) =
\begin{cases}
\mu_{k-1}^*(\gamma),\quad &\text{if } \bar\mu_{k}(\gamma) \geq \mu_{k-1}^*(\gamma), \\
\bar\mu_{k}(\gamma),\quad &\text{if } 0 < \bar\mu_{k}(\gamma) < \mu_{k-1}^*(\gamma) , \\
0,\quad &\text{if } \bar\mu_{k}(\gamma) \leq 0, \\
\end{cases}
\end{equation}
for $k=0,\dots,K$, with $\mu^*_{-1}(\gamma)=1$. Then, the optimal solution of problem (\ref{eq:problemsoft}) is obtained as $\mu_k^*(\gamma^*)$, for $k=1,...,K$, where $\gamma^*$ is the solution of the equality
\begin{equation} \label{eq:gammasol}
\sum_{k=0}^K \mu^*_k(\gamma) A_k = C.
\end{equation}
\end{theorem}
\begin{IEEEproof}
Since the utility function $w$ is increasing, $F_k\geq 0$ and $A_k\geq 0$, we can rewrite the optimization problem of Theorem~\ref{th:soft} as
\begin{align}
\mathrm{min}\ & - \sum_{k=0}^K w(\mu_{k})F_{k}, \\
\text{subject to}\ & \sum_{k=0}^K \mu_{k} A_{k} = C, \label{eq:constr1} \\
& \mu_{k} - \mu_{k-1}\leq 0,\quad k=0,\dots,K, \label{eq:constr2} \\
& -\mu_k \leq 0,\quad k=0,\dots,K, \label{eq:constr3}
\end{align}
with $\mu_{-1}=1$. By introducing dual variables $\gamma$, $\phi_k$ and $\theta_k$ for constraints~\eqref{eq:constr1}, \eqref{eq:constr2} and~\eqref{eq:constr3}, respectively, we see that the KKT optimality conditions are given by
\begin{align}
- w'(\mu_k)F_k + \gamma A_k + \phi_k - \theta_k = 0, & \quad k=0,\dots,K, \label{eq:KKT1} \\
\phi_k (\mu_k - \mu_{k-1}) = 0, & \quad k=0,\dots,K, \label{eq:KKT2} \\
\theta_{k} \mu_k = 0, & \label{eq:KKT3} \\
\phi_k \geq 0, & \quad k=0,\dots,K, \label{eq:KKT4} \\
\theta_k \geq 0, & \quad k=0,\dots,K, \label{eq:KKT5}
\end{align}
as well as~\eqref{eq:constr1}--\eqref{eq:constr3}.

Now, let $\mu_k^*(\gamma)$ satisfy~\eqref{eq:barmusol} and let
\begin{equation}
  \phi_k^* =
  \begin{cases}
  w'(\mu_{k-1}^*(\gamma))F_{k}  - \gamma A_{k}, &\text{if } \bar\mu_0(\gamma)\geq\mu_{k-1}^*(\gamma),\\ 
  0, &\text{if } \bar\mu_{k}(\gamma) < \mu_{k-1}^*(\gamma) 
\end{cases}
\end{equation}
and
\begin{equation}
  \theta_k^* =
  \begin{cases}
  - w'(0)F_{k}  + \gamma A_{k},\quad &\text{if } \bar\mu_{k}(\gamma)\leq 0,\\
  0,\quad &\text{if } \bar\mu_{k}(\gamma) > 0.
  \end{cases}
\end{equation}
Constraints~\eqref{eq:constr2}--\eqref{eq:KKT3} trivially hold. Also, using the fact that $w$ is concave, it is easy to check that~$\mu_k^*(\gamma)$, $\phi_k^*$ and $\theta_k^*$ satisfy constraints~\eqref{eq:KKT4}--\eqref{eq:KKT5} for all $\gamma\geq 0$. It remains to find the value of $\gamma^*$ for which~\eqref{eq:constr1} is satisfied, but this is exactly the condition provided by~\eqref{eq:gammasol}. Finally, note that since $w_i$ is strictly concave and $A_k/F_k>0$ we do not need to consider the case $\lambda<0$.
\end{IEEEproof}

As detailed in Theorem 2, the optimal soft-TTL policy satisfies the condition
\begin{equation}\label{eq:hazard}
w'(\mu_{k}^\mathrm{soft}) \propto \frac{A_k}{F_k} \approx \frac{1}{h(kT)},
\end{equation}
where $h(t) = F'(t)/(1-F(t))$ is the hazard function of the inter-arrival request process, with the prime notation denoting a derivative. The approximate equality in (\ref{eq:hazard}) follows by considering the continuous limit $T\to 0$. The hazard function $h(t)$ measures the probability of observing the next request after $t$ seconds since the most recent request. Using (\ref{eq:hazard}), and recalling that the derivative $w'$ of the concave utility function is non-increasing, we can draw the following conclusions under the typical assumption of a non-increasing hazard function.

First, if the hazard function decreases very quickly, and hence we have clustered or bursty arrivals, then only the fraction $\mu_1$ is non-zero and TTL is optimal. At the other extreme, if the hazard function is constant and hence we have a memoryless arrival process, it is generally optimal to cache a constant fraction of the file and fractional TTL is optimal. In the general intermediate case, soft-TTL is required to obtain the optimal performance.

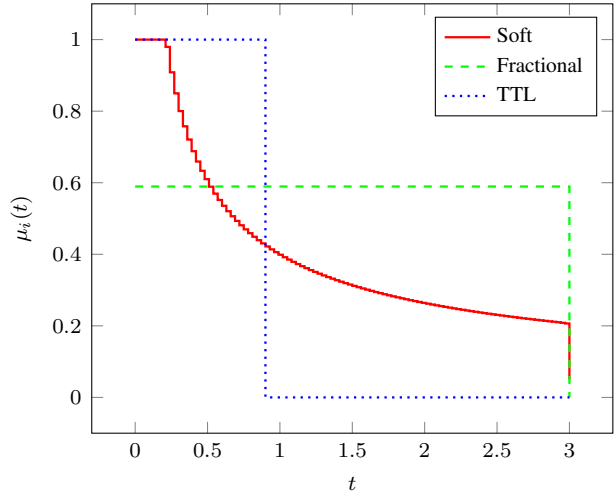
\begin{figure}
\centering
\begin{tikzpicture}
\begin{axis}[
  xlabel=$t$,ylabel={$\mu_i(t)$}, 
  font=\scriptsize,
  legend style={
        cells={anchor=west},
        legend pos=north east,
       font=\scriptsize,
    }
]

\addplot[
  line width=.3mm,color=red, solid,
  const plot mark left,
  ]
table[
  header=false,x index=0,y index=1,
  ]
{singlerun.csv};
\addlegendentry{Soft};
\addplot[
  line width=.3mm,color=green, dashed,
  const plot mark left,
  ]
table[
  header=false,x index=0,y index=2,
  ]
{singlerun.csv};
\addlegendentry{Fractional};
\addplot[
  line width=.3mm,color=blue, dotted,
  const plot mark left,
  ]
table[
  header=false,x index=0,y index=3,
  ]
{singlerun.csv};
\addlegendentry{TTL};
\end{axis}
\end{tikzpicture}
\caption{Optimal policy $\mu_i(t)$ for the TTL, fractional-TTL, and soft-TTL.
\label{fig:profile}}
\end{figure}

\begin{figure}
\centering
\begin{tikzpicture}
\begin{axis}[
  xlabel=$a$,ylabel={$\mathcal{W}$}, 
  font=\scriptsize,
  legend style={
        cells={anchor=west},
        legend pos=north east,
    }
]

\addplot[
  line width=.3mm,color=red, solid,
  ]
table[
  header=false,x index=0,y index=1,
  ]
{shape.csv};
\addlegendentry{Soft};

\addplot[
  line width=.3mm,color=green, dashed,
  ]
table[
  header=false,x index=0,y index=2,
  ]
{shape.csv};
\addlegendentry{Fractional};

\addplot[
  line width=.3mm,color=blue, dotted,
  ]
table[
  header=false,x index=0,y index=3,
  ]
{shape.csv};
\addlegendentry{TTL};

\end{axis}
\end{tikzpicture}
\caption{Optimal utility as a function of Weibull shape parameter $a$.
\label{fig:shape}}
\end{figure}
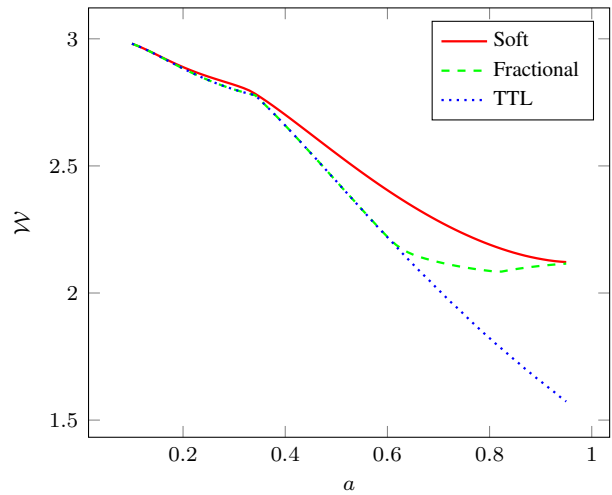

\begin{table}
\centering
\begin{equation*}
\begin{array}{c||cc|cc|cc}
\alpha & \displaystyle W_1^\mathrm{TTL} & \displaystyle W_3^\mathrm{TTL} & \displaystyle W_1^\mathrm{frac} & \displaystyle W_3^\mathrm{frac} & \displaystyle W_1^\mathrm{soft} & \displaystyle W_3^\mathrm{soft} \\
\hline
0 & 0.1963 & 2.8335 & 0.4712 & 2.7913 & 0.3322 & 2.9262  \\
0.5 & 0.4741 & 2.3872 & 0.5667 & 2.4602 & 0.6522 & 2.5391   \\
2 & 0.8204 & 1.6057 & 0.8436 & 1.7578 & 0.8695 & 1.9709 \\
\infty & 0.9215 & 1.3150 & 0.9999 & 0.9999 & 1.0000 & 1.1475
\end{array}
\end{equation*}
\caption{Optimal utility per file as a function of fairness parameter $\alpha$.} \label{table:fairness}
\end{table}

%
%
%
\section{Numerical Results} \label{sec:numerical}
In this section, we present numerical results assuming Weibull distributed inter-arrival times $X_i$ (see, e.g.,~\cite{rausand2004system}). The Weibull distribution has a shape parameter $a_i$ and a scale parameter $b_i$, where the latter can be found as a function of the arrival rate as $b_i = ( \lambda_i\Gamma(1+a_i^{-1}))^{-1}$, where $\Gamma(\cdot)$ is the Gamma function. The shape parameter determines the hazard function $h_i(t) = b_i^{-a_i} a_i t^{a_i-1}$, which is a decreasing function of $t$ for $a_i<1$ and is constant for $a_i=1$. Values of $a_i$ larger than $1$ are not considered, since they yield increasing hazard functions. We consider the same strictly concave utility function $w_i(\mu) = \sqrt{\mu}$ and $a_i=a$ for all files. Unless stated otherwise, we set $J=3$ files; the size of all files to be normalized to $s_i=1$; the arrival rates to be equal to $\lambda_i=1$; the sum-utility obtained with $\alpha=0$; and the cache capacity constraint to be set to $C=1.5$. Furthermore, caching policies (\ref{eq:policy}) are optimized with the parameters $K=100$ and $T=0.03$.

We first illustrate the optimal policies for TTL, fractional-TTL, and soft-TTL when setting $a=0.7$ in Fig.~\ref{fig:profile}. This value of $a$ corresponds to a decreasing hazard function with a non-negligible tail. The solution is shown for one of the files given that, by symmetry, the policies for all files were seen to be the equal. The optimal soft-TTL solution is seen to differ from both TTL and fractional-TTL, and it prescribes a decreasing fraction of cached files over time.

In Fig.~\ref{fig:shape}, we show the sum-utility of the three schemes as a function of the shape parameter $a$. Confirming the insights from the analysis in the previous section, we observe that, when $a$ is very small, and hence the arrivals are clustered, or bursty, TTL is optimal. At the opposite end, when $a=1$ and the inter-arrival processes are memoryless, fractional TTL is optimal and it offers a significant gain over TTL. Finally, for intermediate values of $a$, soft-TTL is to be preferred and it is seen to provide important gains over both TTL and fractional-TTL.

Finally, in Table~\ref{table:fairness} we demonstrate the impact of the fairness parameter $\alpha$ for the case of unequal arrivals. In particular, we set the arrival rates for the three files as $\lambda_1=1$, $\lambda_2=2$ and $\lambda_3=3$, and the other parameters as above. The table provides the optimal values of the utility functions $W_1$ and $W_3$ for the three policies. Beside the proportionality of the utilities in~\eqref{eq:reward} to $\lambda_i$, we observe that, as the fairness parameter $\alpha$ grows larger, soft-TTL enables a large utility $W_1$ of the lower-rate file to be obtained, while also maintaining reasonably high utility $W_3$ of the higher-rate file, whereas fractional-TTL does so at the cost of largely reduced $W_3$.

%
%
%
\bibliographystyle{IEEEtran}
\bibliography{IEEEabrv,soft_eviction_cache}

\begin{thebibliography}{10}
\providecommand{\url}[1]{#1}
\csname url@samestyle\endcsname
\providecommand{\newblock}{\relax}
\providecommand{\bibinfo}[2]{#2}
\providecommand{\BIBentrySTDinterwordspacing}{\spaceskip=0pt\relax}
\providecommand{\BIBentryALTinterwordstretchfactor}{4}
\providecommand{\BIBentryALTinterwordspacing}{\spaceskip=\fontdimen2\font plus
\BIBentryALTinterwordstretchfactor\fontdimen3\font minus
  \fontdimen4\font\relax}
\providecommand{\BIBforeignlanguage}[2]{{%
\expandafter\ifx\csname l@#1\endcsname\relax
\typeout{** WARNING: IEEEtran.bst: No hyphenation pattern has been}%
\typeout{** loaded for the language `#1'. Using the pattern for}%
\typeout{** the default language instead.}%
\else
\language=\csname l@#1\endcsname
\fi
#2}}
\providecommand{\BIBdecl}{\relax}
\BIBdecl

\bibitem{wang2015optimal}
L.~Wang, S.~Bayhan, and J.~Kangasharju, ``Optimal chunking and partial caching
  in information-centric networks,'' \emph{Computer Communications}, vol.~61,
  pp. 48--57, 2015.

\bibitem{sengupta2016cloud}
A.~Sengupta, R.~Tandon, and O.~Simeone, ``Fog-aided wireless networks for
  content delivery: Fundamental latency tradeoffs,'' \emph{IEEE Transactions on
  Information Theory}, vol.~63, no.~10, pp. 6650--6678, 2017.

\bibitem{dehghan2016utility}
M.~Dehghan, L.~Massoulie, D.~Towsley, D.~Menasche, and Y.~C. Tay, ``A utility
  optimization approach to network cache design,'' in \emph{IEEE INFOCOM 2016},
  2016, pp. 1--9.

\bibitem{fofack2012analysis}
N.~C. Fofack, P.~Nain, G.~Neglia, and D.~Towsley, ``Analysis of {TTL}-based
  cache networks,'' in \emph{Performance Evaluation Methodologies and Tools
  (VALUETOOLS), 2012 6th International Conference on}.\hskip 1em plus 0.5em
  minus 0.4em\relax IEEE, 2012, pp. 1--10.

\bibitem{yang2017audience}
Q.~Yang, M.~M. Amiri, and D.~G{\"u}nd{\"u}z, ``Audience retention rate aware
  coded video caching,'' in \emph{Communications Workshops (ICC Workshops),
  2017 IEEE International Conference on}.\hskip 1em plus 0.5em minus
  0.4em\relax IEEE, 2017, pp. 1189--1194.

\bibitem{altman2014distributed}
E.~Altman, K.~Avrachenkov, and J.~Goseling, ``Distributed storage in the
  plane,'' in \emph{Networking Conference, 2014 IFIP}.\hskip 1em plus 0.5em
  minus 0.4em\relax IEEE, 2014, pp. 1--9.

\bibitem{leconte2016placing}
M.~Leconte, G.~Paschos, L.~Gkatzikis, M.~Draief, S.~Vassilaras, and
  S.~Chouvardas, ``Placing dynamic content in caches with small population,''
  in \emph{IEEE INFOCOM 2016}, 2016, pp. 1--9.

\bibitem{feldmann2000characteristics}
A.~Feldmann \emph{et~al.}, ``Characteristics of {TCP} connection arrivals,''
  \emph{Self-similar network traffic and performance evaluation}, pp. 367--399,
  2000.

\bibitem{jung2003modeling}
J.~Jung, A.~W. Berger, and H.~Balakrishnan, ``Modeling {TTL}-based internet
  caches,'' in \emph{IEEE INFOCOM 2003}, 2003, pp. 417--426.

\bibitem{rausand2004system}
M.~Rausand and A.~H{\o}yland, \emph{System reliability theory: models,
  statistical methods, and applications}.\hskip 1em plus 0.5em minus
  0.4em\relax John Wiley \& Sons, 2004.

\bibitem{mo2000fair}
J.~Mo and J.~Walrand, ``Fair end-to-end window-based congestion control,''
  \emph{IEEE/ACM Transactions on networking}, vol.~8, no.~5, pp. 556--567,
  2000.

\bibitem{ross}
S.~M. Ross, \emph{Stochastic Processes}.\hskip 1em plus 0.5em minus 0.4em\relax
  Wiley, 1996.

\bibitem{gurobi}
\BIBentryALTinterwordspacing
{Gurobi Optimization, Inc.}, ``Gurobi optimizer reference manual,'' 2018.
  [Online]. Available: \url{http://www.gurobi.com}
\BIBentrySTDinterwordspacing

\bibitem{boyd2004convex}
S.~Boyd and L.~Vandenberghe, \emph{Convex optimization}.\hskip 1em plus 0.5em
  minus 0.4em\relax Cambridge university press, 2004.

\bibitem{bretthauer1995nonlinear}
K.~M. Bretthauer and B.~Shetty, ``The nonlinear resource allocation problem,''
  \emph{Operations Research}, vol.~43, no.~4, pp. 670--683, 1995.

\end{thebibliography}

\end{document}